# Fast raster scan multiplexed charge stability measurements toward high-throughput quantum dot array calibration

Wonjin Jang,[1†] Min-Kyun Cho, [1†] Myungwon Lee,[1] Changki Hong,[2] Jehyun Kim,[1] Hwanchul Jung,[2]

Yunchul Chung,[2]

Vladimir Umansky,[3] and Dohun Kim[1*]

[1]*Department of Physics and Astronomy, and Institute of Applied Physics, Seoul National University, Seoul 08826, Korea*

[2] *Department of Physics, Pusan National University, Busan 46241, Korea*

[3]*Braun Center for Submicron Research, Department of Condensed Matter Physics, Weizmann Institute of Science, Rehovot 76100, Israel*

[†]*These authors contributed equally to this work,*

*\*Corresponding author: ycchung7@gmail.com, dohunkim@snu.ac.kr*

**Abstract**

We report raster scan multiplexed charge-stability diagram measurements for tuning multiple gate-defined quantum dots in GaAs/AlGaAs heterostructures. We evaluate the charge sensitivity of the quantum point contact (QPC) in both radio frequency (rf)-reflectometry and direct current (dc)-transport modes, where we measure the signal-to-noise ratio (SNR) of 40 for rf-QPC with integration time per pixel of $10 ms$, corresponding to $1.14 ms$ for resolving single electron transition in few electron regime. The high SNR for reasonable integration time allows fast two-dimensional (2D) scanning, which we use to facilitate double and triple quantum dot tuning process. We configure highly stable raster scan multiplexed quantum dot tuning platform using a switching matrix and transformer-coupled alternating current (ac) ramp sources with software control. As an example of high-throughput multiple quantum dot tuning, we demonstrate systematic triple quantum dot (TQD) formation using this platform in which a multiplexed combination of 2D scans enables the identification of few electron regime in multiple quantum dots in just a few minutes. The method presented here is general, and we expect that the tuning platform is applicable to more complex multiple quantum dot arrays, allowing efficient quantum dot system Hamiltonian parameter calibration.

Semiconductor quantum dot systems have emerged as a promising platform for studying quantum information processing [1-4] and simulation of mesoscopic quantum systems [5-8]. Common to the most quantum dot experiments, the measurement of charge stability diagram based on Coulomb blockade phenomena is an indispensable prerequisite step as this measurement provides essential information of the system [9]. Such a mapping process can be taken either via observing conductance through quantum dots [10-12] or using charge sensing techniques [13-15].

Important developments have been made for controlling multiple quantum dot charge and spin states in quantum dot systems [3,16]. Examples include the invention of the exchange-only qubit in triple quantum dots (TQD) [4,17,18] or quantum simulation of the Mott-Hubbard transition in TQD [7] where careful tuning of charge stability diagram provides an essential route to construct the desired experimental platform. Evidently, as the number of quantum dots increases, developments in the efficient system tuning process will be of central importance, because the possible parameter space becomes multi-dimensional.

Moreover, fast calibration of the charge states in the quantum dots is practically important since most quantum dot experiments involves slow drift due to uncontrolled environmental and material factors [19,20]. At least in the coarse tuning process, especially for multiple quantum dot arrays, it is more desirable to focus on obtaining the rough and multiple two-dimensional (2D) charge stability diagrams quickly rather than examining small parameter space with high resolution. This strategy can speed up the calibration process, minimizing the effect of possible drift. Indeed, advances have been made to enhance the data acquisition speed for charge sensing. The radio frequency (rf) sensing technique allows fast and high sensitivity detection of charge states in the dot reaching a few $\mu s$ to resolve single electron transitions [21,22]. More recent advances exploit Josephson parametric amplifier (JPA) [23] to enable video-rate real time tuning of the quantum dot system. These measurements, however, have demonstrated fast charge stability diagram scans up to DQD, and extension to multiple quantum dot systems is still not straightforward.

In this work, we demonstrate fast mapping of the charge-stability diagram of multiple quantum dots within a switched (multiplexed) measurement platform. High throughput raster scan combined with rf-reflectometry achieves fast plotting of large-scale charge-stability diagrams in a few seconds. By applying the switching system to the dot tuning process, fast acquisition of stability diagrams spanned by different sets of gates is allowed. This feature is especially useful for multiple dot calibration, since the stability diagrams by different combinations of gates are required for the characterization of the quantum dot system. We demonstrate the process of tuning TQD with a raster scan multiplexed scheme, where identification of few electron regime in TQD only takes a few minutes without prior knowledge of the dot configuration. The method presented here can be extended to more complex gate architectures, allowing a feasible tuning platform for multiple quantum dot based quantum information and quantum simulation devices.

The device for multiple quantum dots, shown in Fig. 1, consists of a GaAs/AlGaAs heterostructure with a 2D electron gas (2DEG) formed 40 nm below the surface. Quantum dots are defined by depleting the 2DEG using a Ti/Au top gate. Details of the device fabrication are described in Refs. [8,24]. The device was designed to contain up to six quantum dots and four different QPC sensors, As an initial demonstration, we focused on forming upper TQD whose energy is mainly controlled by plunger gates $V_1$, $V_2$, and $V_3$. Each gate was connected to an isolated voltage source (Stanford Research Systems, SIM928) outside the dilution refrigerator through a combination of a copper powder filter and a low-pass resistance-capacitance (RC) filter with cutoff frequency on the order of 3 MHz. The copper powder acted as a high frequency ( > GHz) noise absorber [25].

Charge sensing using the QPC (Fig. 1, upper right sensor) was operated in both near direct current (dc) and rf-reflectometry measurement mode. For near dc measurement, current through the Ohmic contacts was amplified at room temperature by a homemade current preamplifier [26,27], and the signal acquisition was performed with a commercial data acquisition card (National Instruments, NI USB-9215A). We measured differential conductance by modulating $V_2$ voltage at a few kHz with amplitude of 0.8 mV. To perform rf-reflectometry, an impedance matching circuit was made by combining a 2200 nH chip inductor and a parasitic capacitance on the order of 1 pF. An additional 100 pF chip capacitor was connected to the Ohmic contacts in series, acting as an rf ground. Given a reflected signal amplified first by a commercial cryogenic amplifier (Caltech Microwave Research Group, CITLF2) at 4 K, demodulation was performed at room temperature using a high frequency lock-in amplifier (Zurich Instruments, UHFLI) at the carrier frequency around 89 MHz. Similar to dc measurement, we used heterodyne detection scheme [22], where we applied the carrier frequency to an rf ohmic contact and, at the same time, modulated $V_2$ at 1 MHz with amplitude of 0.8 mV.

Figure 1b shows the block diagram for performing multiplexed raster scans. We used a two-channel floating voltage output arbitrary function generator (AFG, Tektronix AFG320) to apply synchronized and successive ac ramp waveforms to selected two different gates, in addition to highly stable dc source that was set to the average confining voltage. We applied the waveform with frequency of 40 Hz to one axis (*x*-axis) and of 1 Hz to the other axis (*y*-axis). For multiplexing, a switching matrix (Keithley, 708A switching system) was placed between the AFG and a homemade transformer-coupled ac + dc box. The floating output capability of the AFG, floating input-output of the switching matrix, and transformer-coupled ac coupling technique minimized possible ground loops. Two input channels of the switching system were connected to the two output channels of the AFG. By selectively switching the combination of outputs of the switching matrix, it was possible to examine a DQD of interest using raster scan while other potential defining gate voltages were set with a highly stable isolated voltage source. Moreover, the switching matrix enabled arbitrary selection of the input-to-output combination.

In our setup, dc lines are heavily low-pass filtered (cut-off frequency of ~1 Hz) for highly stable dc sourcing. While the ac transformer inside the ac + dc box has the advantage of minimizing the ground loop, it can only pass relatively higher frequency signals (> ~30 Hz), allowing only the fast ramp waveform of frequency ~40 Hz and preventing the use of slow ramp waveform of frequency of ~1 Hz with transformer coupling. As shown in Fig. 1b, slow ramp waveforms were directly put to the dc lines via dc combiners before the low-pass filters, and the fast ramp waveforms were put to the ac port of the ac + dc box to be combined with the heavily filtered dc bias. In this manner, both the fast waveform and the slow waveform could be applied to the gates while taking the advantage of highly stable dc sourcing and minimizing the possible ground loops. As can be seen in Fig. 1b, either the fast or the slow ramp waveforms could be put to $V_1$ via the ac + dc box or the combiner, depending on the switching configuration, and only the slow (fast) ramp waveform could be put to $V_2$ ($V_3$). Thus, $V_1 - V_2$ (fast - slow), $V_3 - V_2$ (fast - slow), and $V_3 - V_1$ (fast - slow) stability diagrams could be obtained, which are full combinations of 2D diagrams for TQD characterization. Using this scheme, we expect that the extension to an *N*-quantum dot system would be straightforward; (*N*-2) gates require both the dc combiner and the ac + dc combiner, while the remaining two lines require only the dc combiner or ac + dc combiner, respectively. In this manner any combination of gates for raster scanning would be feasible.

As shown in Fig. 1c, we used a waiting time of about $5ms$ for each *x*-axis (*y*-axis) ramp waveform in order to allow the response of the filtered dc line to follow the abrupt voltage change at the end of each ramp waveform. We applied the sync output of the waveform generator to the trigger input of the lock in amplifier (UHFLI), which we used for signal demodulation, and triggered 2D data acquisition, as well as real-time frame average. Including waiting time, as shown below, we typically observed a clear stability diagram with 50 frame average, so that an averaged 2D charge stability measurement roughly takes on the order of one minute.

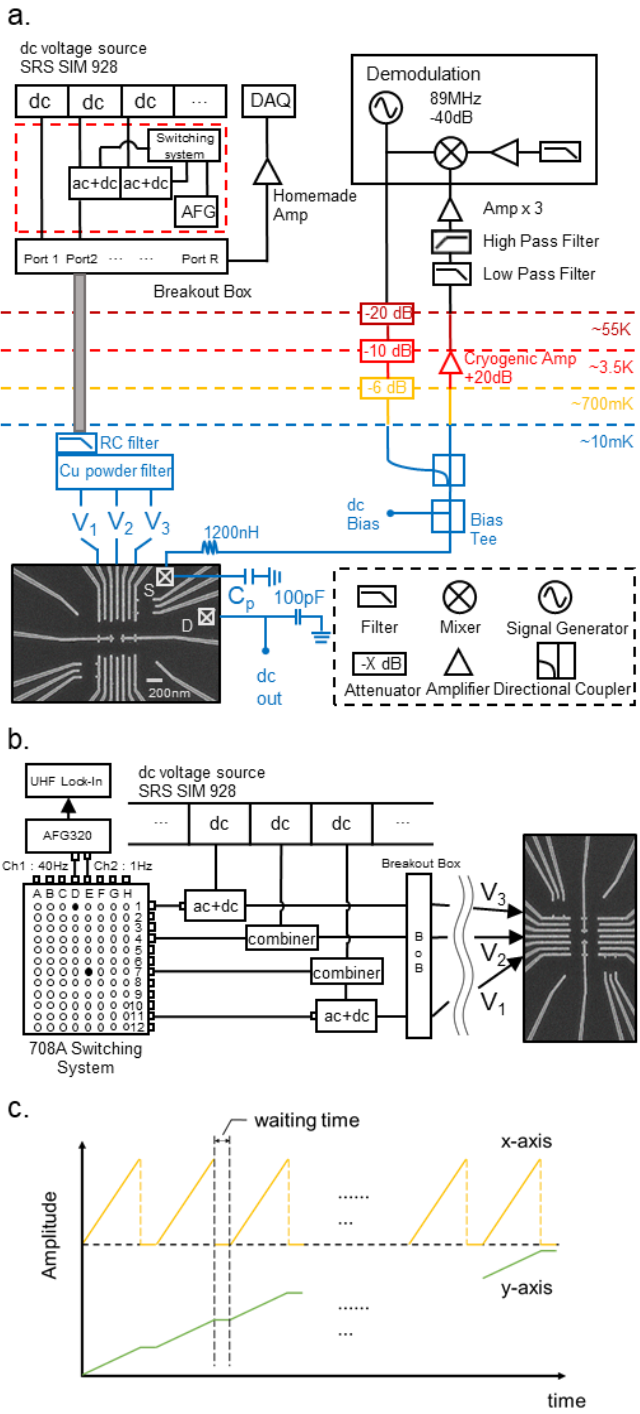

**Figure 1**

The fast scan was possible with the aid of the efficient charge sensing technique [21,22]. We evaluated the signal-to-noise ratio (SNR) of the QPCs in rf-reflectometry and dc transport modes and demonstrated that both modes are, in principle, compatible with the multiplexed raster scan platform. For the comparison of the SNRs on equal footing, we first formed a DQD with the gates $V_2$ and $V_3$, as shown in Figs. 2a and 2b, and focused on the charge transition of the $V_3$ dot for both rf and dc sensing mode. The integration time per pixel for the stability diagram in Figs. 2a and 2b is $10 ms$. The result shows that the scanning method we used does not produce additional systematic drift to the system, which allows systematic investigation of SNR with varying integration time per pixel. The dc data in the Fig. 2b shows quasi-periodic noise of which we ascribe the origin to colored noise peaks including the multiples of 60 Hz line noise due to imperfect elimination of ground loops in our noise spectrum, whose sum and difference frequency with respect to dc-current lock-in modulation frequency can be picked up in the demodulation process. However, we show in the Supplementary Material that these colored noises exist prior introducing the multiplexing setup, and we further show that the multiplexing system does not introduce too much additional noise in our setup. By comparing concurrently acquired rf and dc data, we neglect unphysical features in the present work. Moreover, we investigated the signal at the last charge transition line in order to estimate the lower bound of the sensitivity and demonstrated fast sensing of quantum dots in the few electron regime. SNRs of the sensors were obtained by comparing differential signal change upon one electron transition ($dV_{RF}$, $dV_{DC}$) through gate voltage modulation with background fluctuation. Therefore the SNR is defined by [23,26],

$$SNR = \frac{\Delta V^2}{\sigma^2}, \qquad \text{Eq. (1)}$$

where, as shown in Fig. 2c, $\Delta V$ corresponds to the amplitude of the signal and $\sigma$ corresponds to the root-mean-square amplitude of the background noise in the line cut of the stability diagram. As can be inferred from Eq. 1, SNR tends to increase along the input power of the carrier signal [22]. However, to minimize the power broadening, and to suppress the generation of 2nd harmonics in the signal, we have put rf carrier signal of amplitude -91dBm at the sample. Figure 2d shows the variation of SNR as a function of integration time per pixel, $\tau$. Sensitivity can be obtained from minimum integration time $\tau_{min}$, for which SNR = 1. The SNR non-linearly increases with respect to integration time (Fig. 2d), which is likely due to the colored noise peaks in the current spectrum (see Supplementary Material). We aim to provide a rough estimation of the sensitivity of our sensors by linearly extrapolating the data in the linear SNR vs. $\tau$ regime (Fig. 2d) from which we obtain minimum integration time for rf-reflectometry (dc-charge sensing) $\tau_{min}^{RF} = 1.14 ms$ ($\tau_{min}^{DC} = 2.42 ms$) for resolving one electron transition in the quantum dot in the single electron regime and corresponding sensitivity $S = e\sqrt{\tau_{min}}$ of rf-reflectometry (dc-charge sensing), $S_{RF} = 3.38 \times 10^{-2} e/\sqrt{Hz}$ ($S_{DC} = 4.92 \times 10^{-2} e/\sqrt{Hz}$).

We expect the sensitivity of the rf sensor may be further improved by optimizing the impedance matching condition, and the QPC gate configuration, for example dot-to-sensor distance. However, as shown below, the sensitivity we achieved allows reasonable optimization between measurement time vs. resolution, so that the multiplexed raster scan method is adoptable. We also find that even the sensitivity of the QPC in the dc measurement mode can be used for multiplexed raster scan with several times longer measurement time.

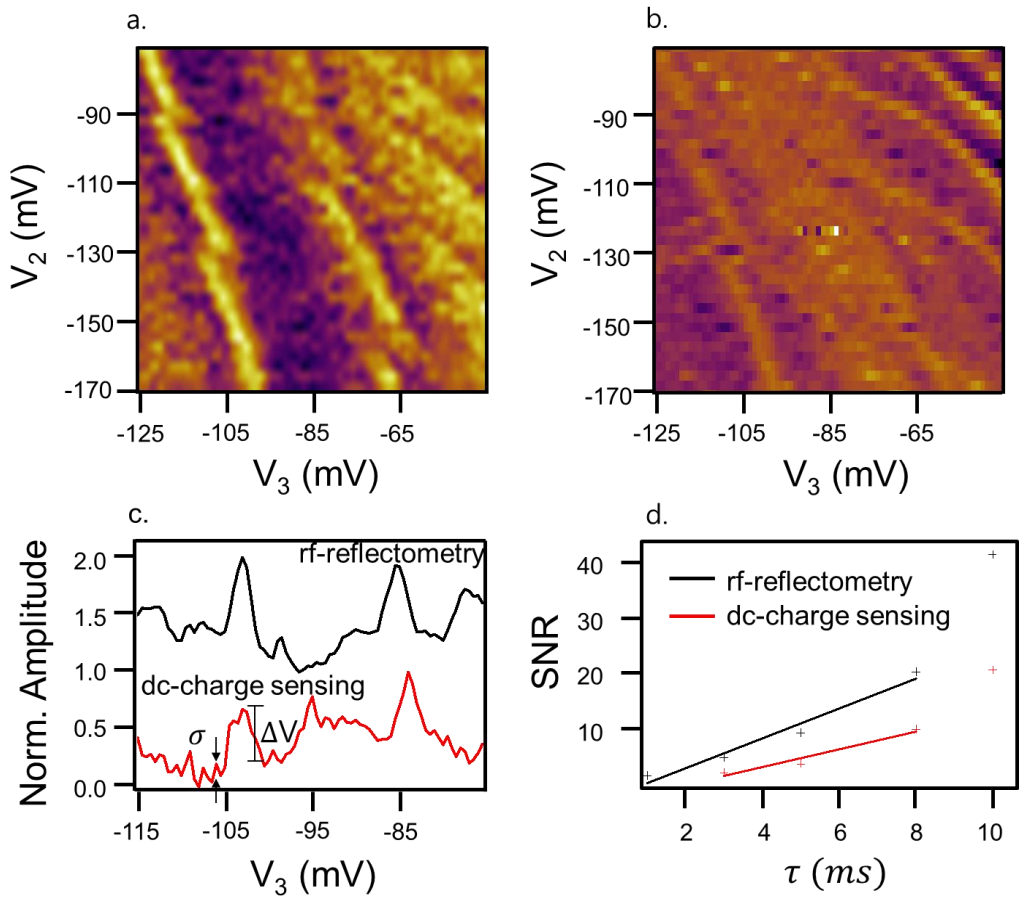

**Figure 2**

As an initial demonstration, we highlight the process of tuning TQD, showing that the enhanced data acquisition rate with multiplexed raster scan can be applied to fast quantum dot array calibration. Figure 3 shows 2D charge stability diagrams by raster scanning sets of gate voltages. In Fig. 3a, a thick horizontal line crosses the diagram and is assumed to be the charging line of the third dot (coupled to $V_3$), which is the closest to the charge sensor. More clearly in Fig. 3c, charge transition lines of three different slopes can be seen, showing evidence of the formation of TQD. Each panel is the average of 30 frames of raster scans, which took roughly one minute. Moreover, selecting sets of scanning voltages was done by the switching matrix, eliminating the need to change the line configuration manually.

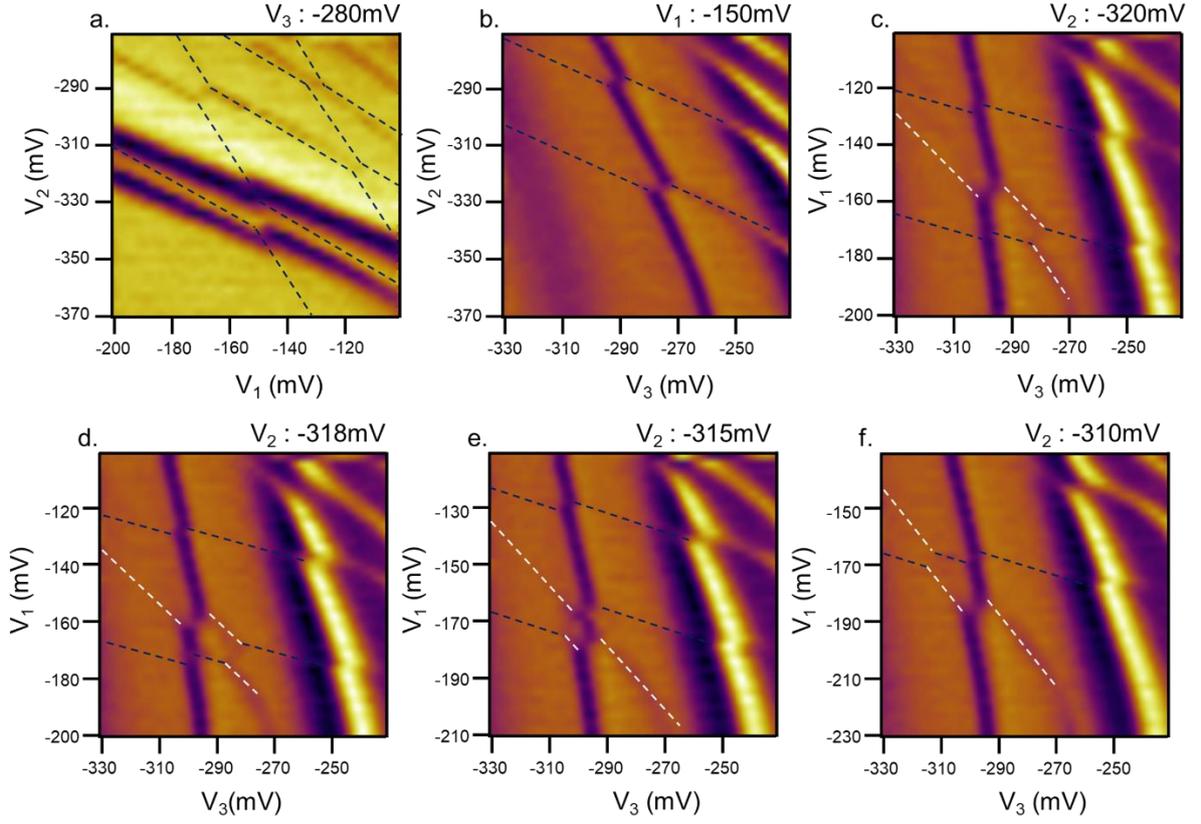

**Figure 3.**

To confirm that the additional slope in Fig. 3c (white dashed line) is the signature of the middle quantum dot, we varied $V_2$ and monitored the response of the signal. As the gate voltage became more positive, the white dashed line systematically moved to more negative voltages, while transition lines coupled to $V_1$ and $V_3$ were relatively unaffected (Fig. 3d ~3f). This shows that this charge transition is the signature of the middle quantum dot, confirming the formation of TQD in the few electron regime. The capability of examining systematic variation of 2D cuts in a short time not only simplified the initial quantum dot array tuning process but also helped to achieve multiple quantum dot array with desired inter-dot interaction strength quickly. Moreover, the observation of nontrivial many-body states with careful tuning of inter-dot interaction vs. on-site energy has become a very important topic in the quantum dot research field [5,8]. For example, in Fig. 3e, we reach the regime where all charge transition lines are crossing each other, near the regime where TQD systems are known to exhibit attractive Coulomb interaction and electron pair tunneling [24,28]. In this sense, the switched (multiplexed) fast scanning offers an efficient route to tune the multi quantum dot array system having desired Hamiltonian parameters. Clearly for increasing system size, the number of 2D scans needed to characterize the possible inter-dot interaction increases quadratically, and we expect that our proposed method can be straightforwardly extended to more complex quantum dot platforms [29,30] and greatly reduce experimental overhead.

The implementation of a floating, ac-coupled ramp source, charge sensing technique and scanning gate multiplexing enables fast measurement of combinations of 2D charge stability diagrams in multiple quantum dot arrays. Charge sensing of quantum dots using QPC in both rf-reflectometry and near dc measurement modes is compatible with our method, with a few times longer integration time for near dc measurement, showing immediate adaptability of the proposed tuning method to the most of the existing quantum dot experiments. As an example of efficient tuning, we have performed fast tuning of TQD starting from nearest neighbor DQDs and systematically achieved the single electron regime in TQD in a few minutes, which is the region of interest for studying quantum dot qubits encoded by multiple spin states [4,17,18,31] or quantum simulation [7,8]. Note that our method only uses RC filtered near dc lines supporting waveforms of frequency up to a few 10 kHz so that rf and microwave lines are available for high frequency qubit manipulations and microwave spectroscopy of the quantum dots [2,32] concurrent with raster scanning. We regard the present method as a compromise between conventional, slow, point-by-point, software controlled scanning and a recently developed technique that is ultrafast but requires a special amplifier [23]. As an intermediate optimization, we have demonstrated that stable ac-coupled multiplexing and conventional charge sensing enable quasi-real time tuning of multiple quantum dots with minimal experimental overhead. It is expected that if our multiplexed tuning scheme is combined with quantum dot systems with more sensitive sensors, even faster tuning of multiple quantum dot array would be possible. As scalability is one of the key features that makes the semiconductor quantum dot an attractive quantum information processing platform [30,33], we expect that our fast tuning process using the switching technique may provide an efficient route for the manipulation of scaled multiple quantum dot systems.

See supplementary material for the comparison of the noise spectrums before and after the switching matrix is applied to the setup.

**Acknowledgements**

  This work was supported by Samsung Science and Technology Foundation under Project Number SSTF-BA1502-03. Cryogenic measurement was performed using instruments partially supported by Research Resettlement Fund for the new faculty of Seoul National University, and Creative-Pioneering Researchers Program through Seoul National University (SNU).

**Figure Captions**

**FIG. 1.** (a) Complete circuit diagram of the experiment, including scanning electron microscopy (SEM) image of a representative device. For clarity, only plunger gates used for raster scanning are denoted in the diagram. Ramp waveforms from the arbitrary function generator (AFG) go through the resistor-capacitor (RC) and powder filtered direct current (dc) lines to three different plunger gates ($V_1$, $V_2$, and $V_3$). Demodulation of the reflected radio frequency (rf) signal is performed at room temperature by mixing the reflected rf signal with the local oscillator, yielding an intermediate frequency signal of 1 MHz (see text for more discussion). (b) Close-up view of the switched measurement system (red dashed box in Fig. 1a). Ramp waveforms are generated by the AFG and put into two different input channels of the switching matrix, making arbitrary combination of gates feasible for multiplexed raster scanning. Line configuration as shown in (b) is made to facilitate the raster scanning under the restrictions of cut-off frequency of the dc lines (~1 Hz), and the alternating current (ac) + dc box (~30 Hz). $V_1$ is capable of either fast (~40 Hz) or slow (~1 Hz) ramping as needed, while $V_2$ and $V_3$ are capable of only slow and fast ramping, respectively. (see main text for detail). (c) Ramp waveforms used for multiplexed two-dimensional (2D) scanning. Wait time of $5ms$ was added to each end of the fast ramp waveform to allow the signal line to respond to the abrupt voltage change. The data acquired during the wait time was discarded.

**FIG. 2.** Signal-to-noise ratio (SNR) measurement. (a), (b) Charge-stability diagram of the double quantum dot (DQD) spanned by $V_2$, and $V_3$. By modulating the $V_2$ gate voltage, charge stability diagram of the DQD with differential radio frequency (rf) (a), and direct current (dc) signals (b) are acquired using $10ms$ of integration time per pixel. (c) Horizontal line-cut of the diagrams in (a) and (b) at $V_2$ = -150 mV. The amplitudes are rescaled to [0, 1] range and offset by 1 for comparison. SNR of the last transition in the third dot (coupled to $V_3$) is obtained from the line-cut according to the Eq. (1). (d) SNR as a function of the effective integration time per pixel. The solid lines are fits to SNR $\propto \tau$, where extrapolation gives the minimum integration time $\tau_{min}^{RF} = 1.14 ms$ and $\tau_{min}^{DC} = 2.42 ms$.

**FIG. 3.** Charge stability diagrams acquired with the multiplexed raster scanning. double quantum dot (DQD) charge stability diagram spanned by (a) $V_1$ - $V_2$, (b) $V_2$ - $V_3$, and (c) $V_1$ - $V_3$. Stability diagram spanned by $V_1$ - $V_3$ with (d) $V_2$ = -318 mV, (e) $V_2$ = -315 mV, and (f) $V_2$ = -310 mV. White dashed lines denote the signal of the dot coupled to $V_2$. As $V_2$ gets more positive, the charging line coupled to $V_2$ systematically moves to more negative voltages with less effect on the lines coupled to $V_1$ and $V_3$. Particularly in (f), all three charging lines are crossing each other, resulting in four triple points and showing 'zig-zag' behavior [28]. Each diagram was obtained in 30 s via differential radio frequency (rf)-reflectometry. Dotted lines are guides to the eye.

**Fast raster scan multiplexed charge stability measurements toward high-throughput quantum dot array calibration**

**Supplementary Note: Noise spectrum analysis**

To confirm that the switching matrix does not introduce additional noise into the measurement system, and to explain the non-linear signal to noise ratio (SNR) vs integration time plot (Fig. 2d), we have analyzed the noise spectrum of the dc-current through quantum point contact before and after utilizing the switching matrix. The quantum point contact current in the device similar to the one used in our manuscript was amplified at room temperature by a homemade battery-operated current preamplifier [1], and low pass filtered (cutoff frequency of 100 kHz) by a low-noise voltage preamplifier (Stanford Research Systems, SR560). The spectrums were measured with the spectrum analyzer module in our lock in amplifier (Zurich Instruments, UHFLI) in resolution bandwidth of 0.42 Hz.

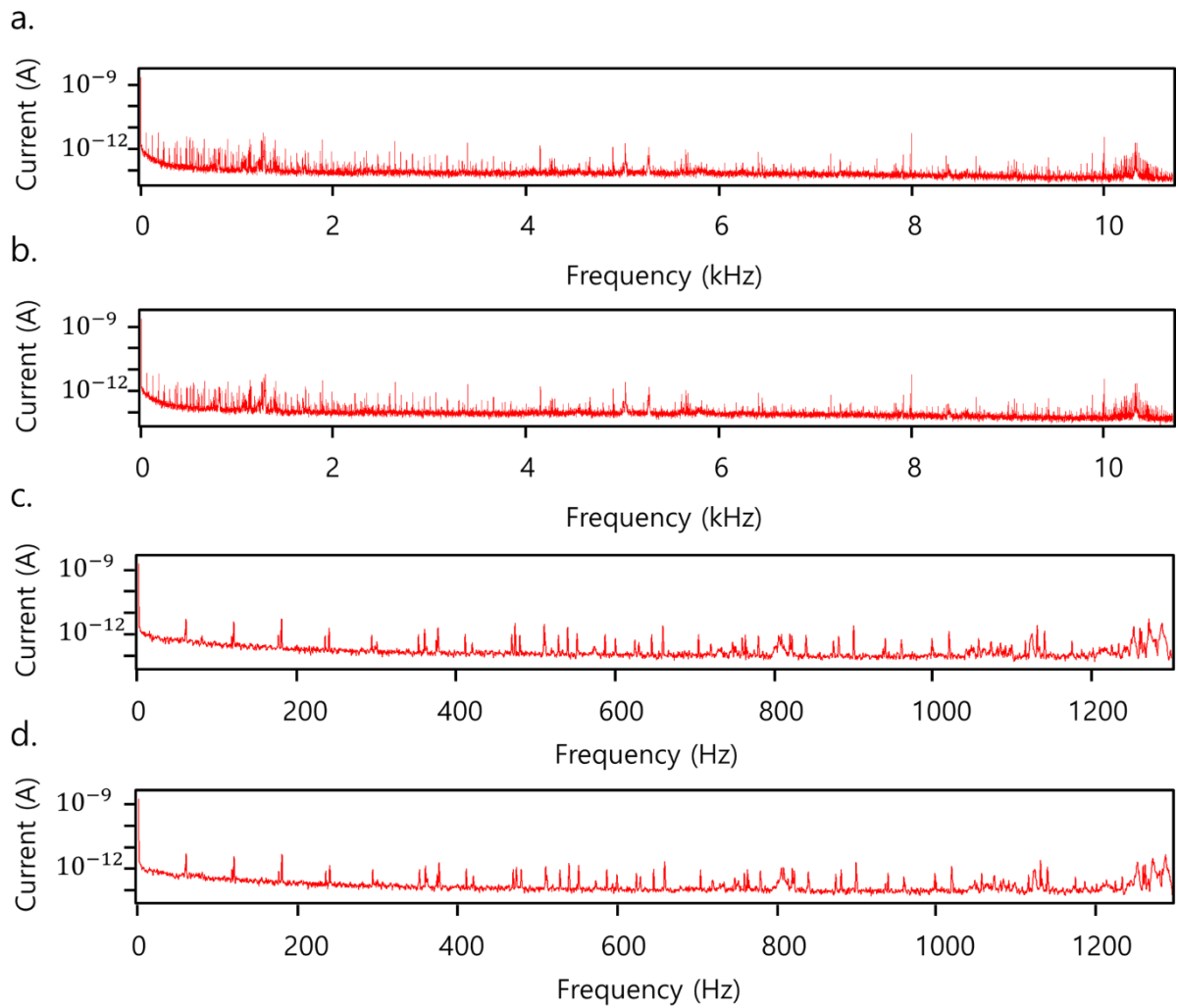

**Supplementary Figure 1.** Noise spectrums of the current through the quantum point contact to ~ 10 kHz frequency range **(a)** with and **(b)** without the multiplexing setup. **(c)** close-up view of **(a)**, and **(d)** close-up view of **(b)** in the dc to ~ 1 kHz frequency range.

There exist 60 Hz line noise and its multiple harmonics both before and after the switching matrix is applied (Supplementary Figure 1c, and 1d). Heavy use of rf-electronics whose ground is typically connected to chassis ground and power supply may account for such noises. Moreover, the colored noise peaks irrelevant to multiple harmonics of 60 Hz are likely caused by triboelectric noise or ground loops occurring inside the fridge. Such noises may be ascribed for the cause of periodic features in Fig. 2 in the manuscript as the sum and difference frequency with respect to dc-current lock-in modulation frequency can be picked up in the demodulation process. Also, we assume that the colored noise peaks may also contribute to the non-linear behavior of the SNR about the integration time (Fig. 2d), as relatively high frequency noise peaks may be limiting the SNR in the short integration time regime.

Importantly, by comparing the Figs. (a,c) and (b,d), it can be inferred that the noise spectrum with the switching matrix applied is not exhibiting significant differences, but only fine features deviate from the case without the switching matrix. This implies that the switching matrix does not induce too much additional noise into the system which also shows that no ground loop is created when the additional instrument is installed. Thereby, utilizing switching matrix for multiplexed quantum dot tuning scheme may be a feasible option for calibrating multiple quantum dot arrays without introducing extra noises into the measurement system.